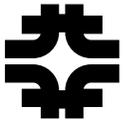

# THEORY AND SUPPRESSION OF MULTIBUNCH BEAM BREAKUP IN LINEAR COLLIDERS*

C. L. Bohn and K.-Y. Ng, Fermilab, Batavia, IL 60510, USA

*Abstract*

We recently developed an analytic theory of cumulative multibunch beam breakup that includes a linear variation of transverse focusing across the bunch train. The focusing variation saturates the exponential growth of the beam breakup and establishes an algebraic decay of the transverse bunch displacement versus bunch number. In this paper we illustrate how the focusing variation works to suppress multibunch beam breakup, as well as how the mechanism scales with accelerator and beam parameters.

## 1 RESULTS OF ANALYTIC THEORY

We recently developed an analytic theory of cumulative multibunch beam breakup (MBBU) with a linear variation of transverse focusing (a form of BNS damping) along the bunch train [1]. It is based on a continuum form of the equation of motion of point bunches in which the discrete transverse kicks imparted by the rf structures are smoothed along the linac. For a generic linac in a linear collider, the solution is:

$$\frac{|\delta x_M(\sigma)|}{x_o} \cong \left[\frac{\gamma(0)}{\gamma(\sigma)}\right]^{1/4} \frac{\sqrt{E}\ \exp[c(\eta)E - (M\omega\tau/2Q)]}{4M\sqrt{2\pi}\ |\sin(\omega\tau/2)|}$$

$$\times \begin{cases} \dfrac{1}{|1-\eta^2|^{1/4}}; & \eta\ \text{not near}\ 1 \\ \left(\dfrac{4}{3}\right)^{1/6}\dfrac{\Gamma(1/3)}{\sqrt{2\pi}}E^{1/6}; & \eta=1 \end{cases},$$

with the auxiliary relations

$$E(\sigma,M) = \Sigma(1)\left[\frac{w_0 q e \ell^2}{2\pi N\gamma(0)mc^2}\frac{\Sigma(\sigma)}{\Sigma(1)}M\right]^{1/2};$$

$$\eta(\sigma,M) = \frac{\pi N}{E(\sigma,M)}|f_\gamma|\frac{\Sigma(\sigma)}{\Sigma(1)}\frac{M}{M};$$

$$c(\eta) = \begin{cases} \dfrac{1}{2}\left[\sqrt{1-\eta^2} + \dfrac{1}{2\eta}\text{atg}\left(\dfrac{2\eta\sqrt{1-\eta^2}}{1-2\eta^2}\right)\right]; & \eta<1 \\ \dfrac{\pi}{4\eta} & \eta\geq 1; \end{cases}$$

$$\Sigma(\sigma) = \int_0^\sigma d\sigma'\left[\frac{\gamma(0)}{\gamma(\sigma')}\right]^{1/2} = \frac{2\sqrt{\gamma(0)}\left[\sqrt{\gamma(\sigma)}-\sqrt{\gamma(0)}\right]}{\gamma(1)-\gamma(0)};$$

$\sigma$ denotes position along the linac normalized against the total linac length $\ell$; $M\omega\tau$ denotes time referenced to the lead bunch -- M is the bunch number, $\omega$ is a representative deflecting dipole-wake frequency, and $\tau$ is the temporal bunch spacing; $|\delta x_M(\sigma)|$ is the envelope bounding the bunch displacement $x_M(\sigma)$ as measured from the steady-state displacement $x_{SS}(\sigma,M\omega\tau)$; $x_0$ is the initial offset of the misaligned input beam; focusing is taken to vary as $\gamma^{-1/2}$; acceleration is taken to be constant; the bunches are presumed to reside neither close to a resonance nor close to a wake zero-crossing. Scaling with respect to the following linac and beam parameters is included: initial and final "energies" $\gamma(0)$, $\gamma(\sigma)$, respectively; deflecting-wake quality factor $Q$; deflecting-wake amplitude $w_0$; bunch charge $q$; number of betatron periods $N$; the total fractional energy spread across the bunch train $f_\gamma$, which is twice the total fractional focusing variation; and the total number of bunches $M$ in a train. The particle charge $e$ and rest-energy $mc^2$ also appear.

A linear focusing variation (or energy spread) across the bunch train may be established by chirping the radiofrequency (rf) power sources, or by using rf-quadrupole focusing magnets. It then influences MBBU through an effective deflecting wake [1],

$$w_{eff}(\sigma,M) = w_0\Theta(M\omega\tau)\exp(-M\omega\tau/2Q_{eff})\sin(M\omega\tau),$$

in which $\Theta(\varsigma)$ is the unit step function and $Q_{eff}$ is an effective quality factor: $(2Q_{eff})^{-1}=(2Q)^{-1}+i\pi N|f_\gamma|/(M\omega\tau)$. To have an impact, the focusing variation needs to be large enough that $|f_\gamma| > M\omega\tau/(2\pi NQ)$. Of course, if $Q$ were sufficiently low, MBBU will be correspondingly low, and the focusing variation would not be needed.

The expression for the MBBU envelope reflects a number of physical processes. The coefficient involving beam energy manifests adiabatic damping. The factor $|\sin(\omega\tau/2)|$ is a relic of a resonance function deriving from coupling between the deflecting-mode frequency and the bunch spacing; the expression is valid only away from wake zero-crossings and resonances. The fractional energy spread $|f_\gamma|$ regulates exponential growth, and finite $Q$ generates exponential damping. The singularity at $\eta=1$ is an unphysical artifact of the solution technique; the solution actually varies smoothly through the value indicated for $\eta=1$. However, exponential growth saturates at $\eta=1$ and afterward, for infinite $Q$, the envelope decays algebraically with bunch number M. Therefore $\eta=1$ is the global maximum in the envelope $|\delta x_M|$, and for the focusing variation to be effective, one should ensure that $\eta=1$ is reached somewhere along the bunch train before it exits the linac, *i.e.*, by ensuring $|f_\gamma| > E(1,M)/(\pi N)$.

---
*Work supported by the Universities Research Association, Inc., under contract DE-AC02-76CH00300 with the U.S. Department of Energy.

## 2 NUMERICAL EXAMPLES

The analytic solution allows one to decipher the linear-collider parameter space in terms of, e.g., the projected emittance, as is done in Ref. [1]. Herein, using numerical examples, we illustrate the aforementioned physical processes associated with a linear focusing variation. Table 1 gives baseline parameters used for this purpose.

Table 1: Baseline Parameters

| Parameter | Value |
|---|---|
| Total initial energy $\gamma(0)mc^2$ | 10 GeV |
| Total final energy $\gamma(1)mc^2$ | 1 TeV |
| Linac length $\ell$ | 10 km |
| Number of betatron periods $N$ | 100 |
| Bunch charge $q$ | 1 nC |
| Number of bunches in train $M$ | 90 |
| Bunch spacing $\tau$ | 2.8 ns |
| Deflecting-wake frequency $\omega/2\pi$ | 14.95 GHz |
| Deflecting-wake quality factor $Q$ | $\infty$ |
| Deflecting-wake amplitude $w_0$ | $10^{15}$ VC$^{-1}$m$^{-2}$ |

### 2.1 Analytic vs. Numerical Solutions

Figure 1, depicting the bunch train at the linac exit, shows good agreement between the envelope $|\delta x_M(1)|$ calculated analytically and bunch displacements $\delta x_M(1)$ calculated by solving the equation of motion numerically. It also shows the qualitative difference in the bunch-train pattern between $\eta<1$ and $\eta>1$; with the parameters of Table 1, $|f_\gamma| = 2.2\%$ corresponds to $\eta=1$. Thus, a modest (few-percent) focusing variation suffices to suppress MBBU, as Stupakov observed in simulations of a contemporary Next Linear Collider lattice [2]. Of course, were the wake amplitude too large or the focusing too weak, then a correspondingly larger focusing/energy spread is required, as is illustrated in Fig. 2.

### 2.2 Saturation of Exponential Growth

As shown in Fig. 3, the difference in the patterns of Fig. 1 arises from saturation of the growth factor $c(\eta)E$. As $\eta$ exceeds unity, the envelope $|\delta x_M|$ decays algebraically, varying as a negative power of M. The gradual decay of MBBU for $\eta>1$ is seen in the bottom curve of Fig. 1. Thus, in suppressing MBBU, a linear focusing variation acts differently from exponential decay that accompanies a finite deflecting-wake $Q$.

The bunch train tends to be centered about the steady-state displacement which, in the presence of a focusing variation, oscillates harmonically with bunch number M. Consequently, as shown in Fig. 4, the bunch train itself assumes a complicated form. Because a linear collider brings bunch trains from two distinct linacs into collision, the final-focus system must damp the displacements to ensure the bunch-to-bunch overlap at the interaction point is sufficient to achieve the desired multibunch luminosity. However, Fig. 4 applies to the case $Q\to\infty$; a low $Q$ would simply leave a residual oscillation from any focusing variation present; the spread of bunches about the steady-state curve would be exponentially damped.

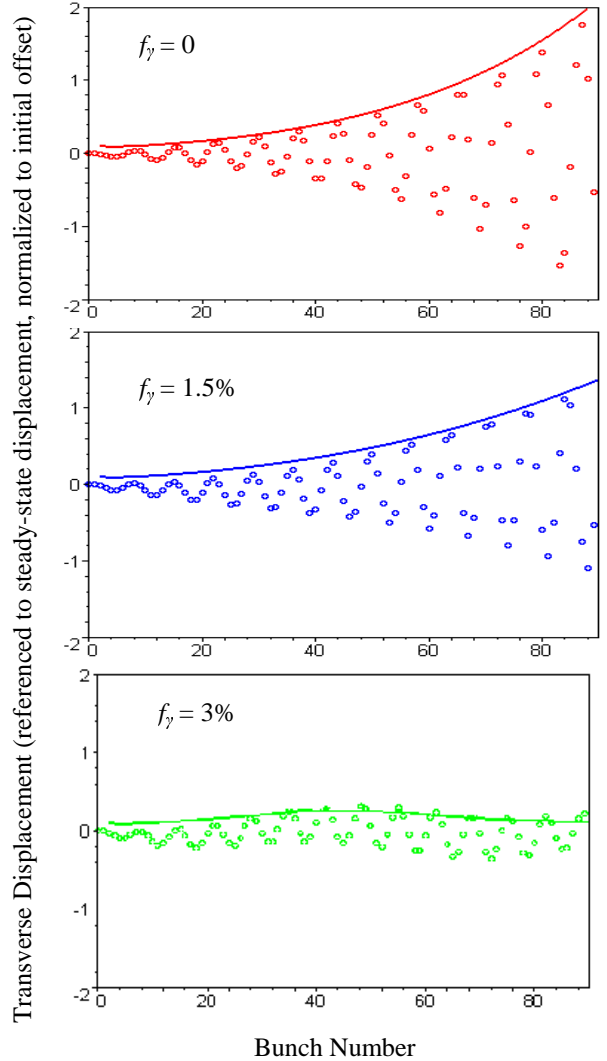

Figure 1: Displacement $(x_M-x_{SS})/x_0$ vs. bunch number M at linac exit for Table 1 parameters and $f_\gamma = 0$ (top), 1.5% (middle), 3% (bottom). Solid curves are analytic solutions for the envelope; circles are numerically calculated displacements.

### 2.3 Finite Deflecting-Mode Quality Factor

Figure 5 depicts the influence of a finite $Q$ relative to that of a nonzero energy spread $|f_\gamma|$. It shows the displacement $|x_{90}-x_{SS}|/x_0$ of the last bunch M=M=90 at the linac exit plotted for various values of $Q$. Given the parameters in Table 1, the energy spread will be useful in suppressing MBBU provided $|f_\gamma(\%)| > 100M\omega\tau/(2\pi NQ) =$

3,800/$Q$. One can see from Fig. 5 how this criterion manifests itself; the displacement is approximately independent of energy spread until the stated threshold is exceeded, after which the displacment drops off relatively fast with increasing $|f_\gamma|$. However, Fig. 5 also shows that the displacement is sensitively dependent on $Q$. Accordingly, designing rf structures for a linear collider involves trading between low deflecting-wake $Q$ and high shunt impedance of the accelerating mode [3].

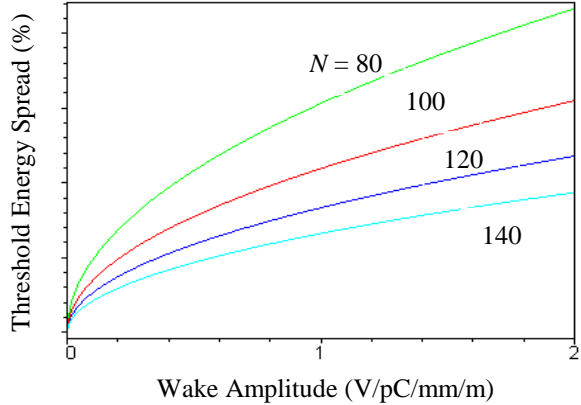

Figure 2: Threshold energy spread $|f_\gamma|$ (corresponding to $\eta=1$ at the linac exit) versus deflecting-wake amplitude $w_0$ and number of betatron periods $N = 80$ (top), 100, 120, 140 (bottom).

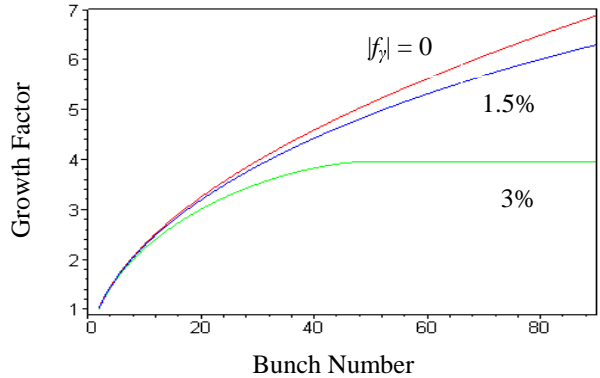

Figure 3: Growth factor $c(\eta)E$ at the linac exit versus bunch number M for $|f_\gamma| = 0$ (top), 1.5% (middle), 3% (bottom).

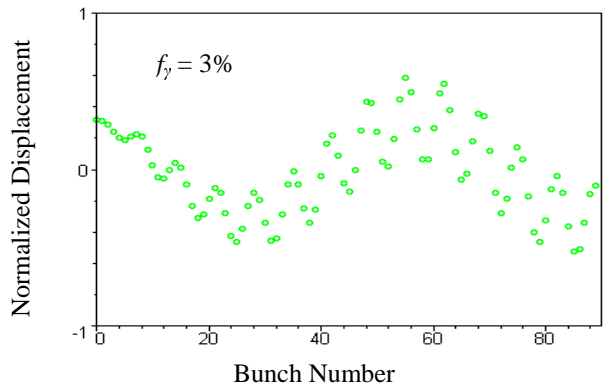

Figure 4: Numerically calculated displacement $x_M/x_0$ at the linac exit versus bunch number M for $f_\gamma = 3\%$.

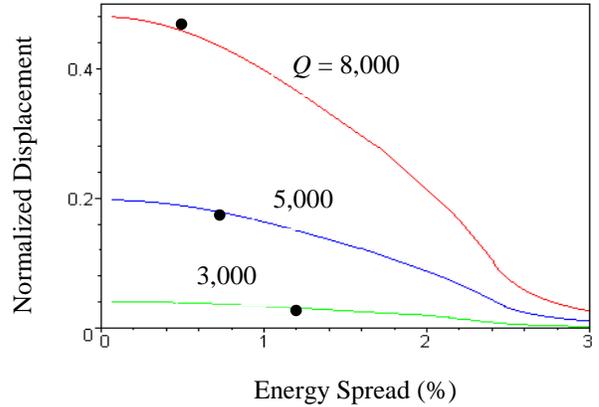

Figure 5: Displacement $|x_{90}-x_{SS}|/x_0$ of the last bunch M=$M$=90 at the linac exit versus energy spread $|f_\gamma|$ for $Q = 8,000$ (top), 5,000 (middle), 3,000 (bottom). The "effectiveness criterion" is $|f_\gamma(\%)| > 3,800/Q$, with the threshold shown for each case (dots).

## 3 CONCLUSIONS

We discussed an analytic solution of the equation of transverse motion for multibunch beam breakup with a linear focusing/energy variation across the bunch train. The solution is, by design, applicable to the main linacs of linear colliders. It constitutes a nontrivial extrapolation from work done in the early 1990s wherein analytic results were derived for all regions of linac parameter space, but without such BNS damping [4].

A key reward is the ability to decipher the inherent parametric scaling. We presented two conditions that both need to be fulfilled for the focusing/energy variation to be effective, one relating to linac and beam parameters separate from the deflecting-wake $Q$ (the "$\eta>1$ criterion"), and the other relating to $Q$ explicitly. With parameters representative of a linear collider, a modest energy spread suffices to suppress MBBU, a finding that is consistent with Stupakov's simulations of an NLC main linac [2]. Of course, the focusing/energy variation cannot be arbitrarily large; practical limitations such as longitudinal beam requirements at the interaction point, lattice chromaticity, etc., impose constraints beyond the two that we described.

The authors are grateful to M. Syphers and G. Stupakov for stimulating discussions.

## REFERENCES

[1] C. L. Bohn, K.-Y. Ng, *Phys. Rev. Lett.* **85**, 984 (2000).
[2] G. Stupakov, SLAC Report No. LCC-0027 (1999), and in these Proceedings.
[3] Z. Li, T. Raubenheimer, these Proceedings.
[4] C. L. Bohn, J. R. Delayen, *Phys. Rev. A* 4**5**, 5964 (1992).